\documentclass[12pt]{article}
\usepackage{bbm}
\usepackage{yfonts}
\pdfoutput=1
\usepackage{amsfonts}
\usepackage[bbgreekl]{mathbbol}
\makeatletter
\def\amsbb{\use@mathgroup \M@U \symAMSb}
\makeatother
\usepackage[bbgreekl]{mathbbol}
\usepackage{amssymb,amsfonts}
\usepackage{graphicx}
\usepackage[top=2.75cm, bottom=2.75cm, left=2.75cm, right=2.75cm]{geometry} 
\usepackage{appendix}

\newcommand{\ep}[1]{\epsilon^{#1}}
\newcommand{\epb}[1]{\bar{\epsilon}^{\,#1}}

\newcommand{\be}{\begin{equation}}
\newcommand{\ee}{\end{equation}}
\newcommand{\ber}{\begin{eqnarray}}
\newcommand{\eer}{\end{eqnarray}}

\newcommand{\Jp}{J^{\mbox{\tiny$(+)$}}}
\newcommand{\Jm}{J^{\mbox{\tiny$(-)$}}}
\newcommand{\Jpm}{J^{\mbox{\tiny$(\pm)$}}}

\def\+{{+\!\!\!+}} 
\def\pp{\mbox{\tiny${}_{\stackrel\+ =}$}}

\newcommand{\nn}{\nonumber}
\newcommand{\pa}{\partial}

\newcommand{\half}{{\textstyle{\frac12}}}
\newcommand{\ihalf}{{\textstyle{\frac i 2}}}
\newcommand{\quart}{{\textstyle{\frac14}}}

\newcommand{\bbX}[1]{\mathbb{X}^{#1}}

\newcommand{\bbD}[1]{\mathbb{D}_{#1}}
\newcommand{\bbDB}[1]{\bar{\mathbb{D}}_{#1}}

\newcommand{\re}[1] {(\ref{#1})}
\newcommand{\nll}{N\!=\!(1,1)}
\newcommand{\nZZ}{N\!=\!(2,2)}
\newcommand{\nff}{N\!=\!(4,4)}
\begin{document}
\begin{titlepage}
\begin{flushright} \small
UUITP-17/14\\
\end{flushright}
\smallskip
\begin{center} 
\LARGE
{\bf  Extended supersymmetry of semichiral sigma model in $4D$.} \\[30mm] 
\large
{\bf}~~{\bf Ulf~Lindstr\"om$^{a}$} \\[20mm]
{ \small\it
$^a$Department of  Physics and Astronomy, Division of Theoretical Physics,
Uppsala University, \\ Box 516, SE-751 20 Uppsala, Sweden \\}
\end{center}

\vspace{10mm}
\centerline{\bfseries Abstract} 
\bigskip

\begin{flushleft}
{\em Briefly}: Using a novel $(1,1)$ superspace formulation of semichiral sigma models with $4D$ target space,  we investigate if  an extended supersymmetry in terms of semichirals is compatible with having a $4D$ target space with torsion.\\
{\em In more detail}:
Semichiral sigma models have $(2,2)$ supersymmetry and Generalized K\"ahler target space geometry by construction. They can also support $(4,4)$ supersymmetry and Generalized Hyperk\"ahler geometry, but when the target space is four dimensional indications are that the geometry is restricted to  Hyperk\"ahler. To investigate this further, we reduce the model to $(1,1)$ superspace and construct the extra (on-shell) supersymmetries there.
We then find the conditions for a lift to $(2,2)$ super space and semichiral fields to exist. Those conditions are shown to hold for  Hyperk\"ahler geometries. The $SU(2)\otimes U(1)$ WZW model, which has $(4,4)$ supersymmetry and  a semichiral description, is also investigated. The additional supersymmetries are found in $(1,1)$ superspace but shown {\em not} to be liftable to a $(2,2)$ semichiral formulation.
\end{flushleft}

\vfill
\end{titlepage}

\renewcommand{\theequation}{\thesection.\arabic{equation}}
\tableofcontents
\section{Introduction}

Generalized K\"ahler geometry is efficiently probed by $(2,2)$  supersymmetric sigma models in $D=2$,   \cite{Lindstrom:2005zr}.  Of particular interest for the present investigation is the symplectic case, i.e., sigma models that depend on semichiral superfields only. Additional supersymmetries for these models were discussed in \cite{Goteman:2010sf}, and in \cite{Goteman:2012qk}. In the latter article focus is on four-dimensional target spaces and it is shown that a very general ansatz for additional supersymmetries leads to an on-shell extended supersymmetry and restricts the target space geometry to be hyperk\"ahler.

In \cite{Goteman:2012qk} this is seen as a shortcoming of the ansatz, since it is argued that the $SU(2)\otimes U(1)$ WZW model of \cite{Rocek:1991vk} constitutes a counterexample. It has nonzero torsion and when coordinatized by chiral and twisted chiral superfields it has ``manifest'' $(4,4)$ supersymmetry. It further has a dual semichiral  description  \cite{LRRUZ} which is then expected to also display the $(4,4)$ supersymmetry\footnote{By ``manifest'' we shall mean ``as realised by transformations of $(2,2)$ superfields.''}.

In this paper we investigate  the possibility that the $(2,2)$ semichiral conditions are incompatible  with ``manifest'' $(4,4)$ transformations. \footnote{Another case of supersymmetries being obstructed occurs when dualisation is along isometries that do not commute with the extra super symmetries. This leads to nonlocal realisations of the extra susys  in the dual model \cite{Bakas:1995hc}. Here, however, the extra susys commute with the isometry used in dualising.}
To study this problem, we descend to $(1,1)$ superspace and develop an on-shell formalism for the extra super symmetries, a formulation which retains the relation to $(2,2)$ semichirals. 
We test this $(1,1)$ formalism on the second supersymmetry (which is non-manifest in $(1,1)$) and then apply it to a hyperk\"ahler geometry which is shown to satisfy the conditions for having a $(2,2)$ semichiral realisation, as expected from \cite{Goteman:2012qk}.

We also derive the extra supersymmetries for the WZW model \cite{Rocek:1991vk}  in $(1,1)$ superspace in the relevant coordinates.
When subjected to the same test they fail to satisfy some of the conditions. This leads to the surprising conclusion that $(4,4)$ supersymmetry in a $(1,1)$ formulation of a $(2,2)$ sigma model with on-shell supersymmetry is incompatible with the introduction of  the $(2,2)$ auxiliary fields.

\section{Background}
\subsection{Semichiral sigma models}
\label{Semichiral sigma models}
Consider a generalized K\"ahler potential \cite{Lindstrom:2005zr}
with one left- and one right semichiral field and their complex conjugates, $K(\bbX{L}, \bbX{R})$, where $L=(\ell, \bar \ell)$ and $R=(r, \bar r)$. The action,
\ber
	S=\int d^2x d^2\theta d^2\bar{\theta}K(\bbX{L}, \bbX{R})
\eer
has manifest $\nZZ$ supersymmetry. The supersymmetry algebra is defined in terms of the anti-commutator of the covariant supersymmetry derivatives as
\ber
\{\bbD{\pm},\bbDB{\pm}\}=i\pa{\pp}
\eer
and the semichiral fields are defined by their chirality constraints as \cite{Buscher:1987uw}
\ber
	\bbDB{+}\bbX{\ell} = 0~, \quad \bbDB{-} \bbX{r} = 0~.
\eer

The geometry of the model is bi-hermitean \cite{Gates:1984nk}, \cite{Buscher:1987uw},  governed by two complex structures $\Jp$ and $\Jm$ that both preserve the metric $\amsbb{G}$  
\ber
\Jpm{}^{t}\amsbb{G}\Jpm ={\amsbb{G}}
\eer
as well as by an anti-symmetric $B$-field whose field strength $H$ enters in the form of torsion in the covariant constancy conditions
\ber
0=\nabla^{(\pm)}\Jpm=\left(\partial+\Gamma^{(0)}\pm\half H\amsbb{G}^{-1}\right)\Jpm~,
\eer
where $\Gamma^{(0)}$ is the Levi-Civita connection. These conditions identify the geometry as bi-hermitean \cite{Gates:1984nk}, or generalized K\"ahler geometry (GKG) \cite{Gualtieri:2003dx}. 

The fact that our superfields are semichiral specifies the  GKG as being of symplectic type where the metric $g$ and the $\amsbb{B}$-field take the form\footnote{This gives the $\amsbb{B}$ field in a particular global gauge as $\amsbb{B}=\amsbb{B}^{(2,0)}+\amsbb{B}^{(0,2)}$ with respect to both complex structures.}
\ber\nn
\amsbb{G}&=&\Omega [\Jp, \Jm]\\
\amsbb{B}&=&\Omega \{\Jp, \Jm \}~.
\eer
The matrix $\Omega$ is defined as 
\ber
\Omega=\half\left(\begin{array}{cc}
0&K_{LR}\cr
-K_{RL}&0\end{array}\right)
\eer
and the submatrix $K_{LR}$ is the Hessian
\ber
K_{LR}=\left( \begin{array}{cc}K_{\ell r}&K_{\ell\bar r}\cr
K_{\bar \ell r} &K_{\bar \ell\bar r}\end{array}\right)~.
\eer

An additional condition results from the target space being four-dimensional and reads  \cite{Buscher:1987uw}
\ber
 \{\Jp, \Jm\}=2 c,~~~~\Rightarrow \amsbb{B}=2 c \,\Omega~,
\eer
where $c$ in general is a function of the coordinates.

\subsection{Extra SUSY}
In \cite{Goteman:2012qk} it is shown that a general ansatz for $(4,4)$ susy in a semichiral sigma model 
\ber\nn
\delta \bbX{\ell} &=&\bar\epsilon^+ \bbDB{+}f(\bbX{L}, \bbX{R}) + g(\bbX{\ell})\bar\epsilon^{-}\bbDB{-} \bbX{\ell}\, + h(\bbX{\ell})\epsilon ^- \bbD{-}\bbX{\ell}~,\\
\delta \bbX{\bar\ell} &=& \ep{+}\bbD{+} \bar f(\bbX{L}, \bbX{R}) + \bar g(\bbX{\bar \ell})\ep{-}\bbD{-}\bbX{\bar\ell}+\bar h(\bbX{\bar \ell}) \epb{-}\bbDB{-}\bbX{\bar \ell}~,\nn\\
\delta \bbX{r}&=& \bar\epsilon^- \bbDB{-}\tilde f(\bbX{L}, \bbX{R}) +\tilde g(\bbX{r})\bar\epsilon^+ \bbDB{+}\bbX{r}+\tilde h(\bbX{r})\epsilon^+ \bbD{+}\bbX{r}~,\nn\\
\delta\bbX{\bar r} &=& \ep{-}\bbD{-} \bar{\tilde f}(\bbX{L}, \bbX{R}) + \bar{\tilde g}(\bbX{\bar r})\ep{+}\bbD{+}\bbX{\bar r}+\bar{\tilde h}(\bbX{\bar r}) \epb{+}\bbDB{+}\bbX{\bar r}~,
\eer
leads to invariance and closure of the algebra only on-shell and provided that the geometry is hyperk\"ahler. The on-shell requirement follows from
\ber\label{c}
[\delta_1,\delta_2]\ell =-\epsilon^+_{[2}\bar\epsilon^+_{1]}|f_{\bar \ell}|^2\partial_\+....
\eer
which has the wrong sign for supersymmetry. It is an interesting fact that on-shell closure of the algebra, together with conditions that come from invariance of the action, requires  that the function $c(\bbX{L}, \bbX{R})$ defined by \re{c}
is constant with absolute value less than one, which means that the geometry is hyperk\"ahler. This on-shell closure is different than the one which arises in the general $(1,1)$ discussion of extended susy \cite{Gates:1984nk} which locates the non-closure of the algebra to the $(+,-)$ sector where the commutator $[J^{(+)},J^{(-)}]$ multiplies the field equation. In the present case left or right susy alone require field equations.

In \cite{Goteman:2012qk} we argue that the ansatz for the additional supersymmetry is too restrictive and should include central charge transformations. The reason for trying to find a more general ansatz is that there is a known example of a BiLP\footnote{This acronym stands for ``bihermitean local product'' and refers to a $2D$ sigma model with chiral and twisted chiral superfields only.} with   $(4,4)$ supersymmetry, mentioned in the introduction, that has a dual semichiral formulation which manifestly violates the hyperk\"ahler condition \cite{LRRUZ}. The duality and extra supersymmetry will be discussed in detail in Sec.4 below. Since the isometry used in the dualisation commutes with the extra supersymmetry, the dual model is expected to have the extra symmetry as well. To investigate this, bearing in mind that the relevant algebra only closes on-shell, we now develop a novel $\nll$ form of the semichiral model and its additional symmetries.

\section{$\nff$ in $\nll$ superspace}
We want to find out under what conditions a semichiral  sigma model in $4D$ supports  additional complex structures
forming an $SU(2)$ algebra\footnote{Corresponding to $(+)$-supersymmetries. The general case also involves 
$(-)$-supersymmetries.}
\ber\label{su2}
I_{(+)}^{(\textfrak a)}I_{(+)}^{(\textfrak b)}=-\delta^{\textfrak a\textfrak b}+\epsilon^{{\textfrak a \textfrak b \textfrak c}}I_{(+)}^{({\textfrak c})}
\eer
with $\textfrak a=1,2,3$ and the identification $I_{(+)}^{(3)}:=J^{(+)}$.
To this end, we discuss the situation in $(1,1)$ superspace \cite{Gates:1984nk}. This discussion is general, only the later applications in the example section will be limited to $4D$. We replace  spinor derivatives according to
\ber
\bbDB{\pm}\to D_\pm+iQ_\pm~,~~ \bbD{\pm}\to D_\pm-iQ_\pm~.
\eer
The general form of the $(2,2)$ sigma model  reduced to $(1,1)$ has a Lagrangian that reads 
\ber\label{completed}
{\cal L} = D_+X^ A\amsbb{E}_{AB}(X)D_-X^C+\Psi_+^RK_{RL}\Psi_-^L := {\cal L}_1+{\cal L}_2~,
\eer
where $\amsbb{E}_{AB}:=\amsbb{G}_{AB}+\amsbb{B}_{AB}$ and we have completed the square  for the spinor auxiliary fields $\psi_\pm$ and defined
\ber\nn\label{bigpsi}
&&\Psi_+^R:=\psi_+^R-D_+X^AJ_{(+)A}^R\\[1mm]
&&\Psi_-^L:=\psi_-^L-J_{(-)A}^LD_-X^A~.
\eer
Assume that we have found the additional transformations of the $(1,1)$ coordinates generated by the $SU(2)$ set of complex structures $I_{(+)}^{(\textfrak a)}$ as in \re{su2}. These transformations leave the ${\cal L}_1$ part of the action invariant. We would now like to extend them to symmetries of the full action and subsequently check if the full set can come from transformations of the $(2,2)$ semichiral fields. 

There are obvious  symmetries we can write down, field equation symmetries (also called Zilch symmetries), but there are many possibilites. 
\subsection{$\nZZ$}

To get a guide to the correct form, we use the fact that we know that one of the symmetries has the correct properties; the one generated by $I_{(+)}^{(3)}=J_{(+)}$.

The full action is invariant under the following transformations:
\ber\nn\label{j3sym}
&&\delta X^L=\epsilon^+JD_+X^L\\[1mm]\nn
&&\delta X^R=\epsilon^+\psi_+^R\\[1mm]\nn
&&\delta \psi^L_-=\epsilon^+JD_+\psi^L_-\\[1mm]
&&\delta \psi^R_+=\epsilon^+D_+^2X^R~,
\eer
where $J$ is the canonical complex structure $diag(i,-i)$.
They give
\ber\nn\label{fullvar}
\delta {\cal L} &&=-2\delta X^ A\nabla_+^{(-)}D_-X^B\amsbb{G}_{AB}+\delta\Psi_+^RK_{RL}\Psi_-^L+\Psi_+^R\delta K_{RL}\Psi_-^L+\Psi_+^RK_{RL}\delta\Psi_-^L\\[1mm]\nn
&&=-2\epsilon^+J_{(+)C}^AD_+X^C\nabla_+^{(-)}D_-X^B\amsbb{G}_{AB}+\Psi_+^R\left(K_{RL}\delta\Psi_-^L+2\epsilon^+\nabla_+^{(-)}D_-X^B\amsbb{G}_{RB}\right)\\[1mm]
&&+\delta\Psi_+^RK_{RL}\Psi_-^L+\Psi_+^R\delta K_{RL}\Psi_-^L~,
\eer
where capital letters from the beginning of the alphabet takes on all values $L$ and $R$.
The first of the terms in the last line is the variation of ${\cal L}_1$ under the $J_{(+)}^{(3)}$ symmetry and vanishes in the action. We evaluate the remaining terms using
\ber\nn
&&\delta \Psi^L_-= \epsilon^+JD_+\Psi^{L}_-
+\epsilon^+J^L_{(-)R'}D_-\Psi^{R'}_+-\epsilon^+J^L_{(-)AR'}\Psi^{R'}_+D_-X^A\\[1mm]\nn
&&~~~~~~~~~+\epsilon^+\left[J_{(+)},J_{(-)}\right]^L_{~A}\left(D_+D_-X^A +D_+X^B\Gamma_{BC}^{(-)~A}D_-X^C\right)\\[1mm]
&&\delta \Psi^R_+= \epsilon^+J^R_{(+)R'}D_+\Psi^{R'}_+-\epsilon^+J^R_{(+)AR'}\Psi^{R'}_+D_+X^A
\eer
which follows from \re{j3sym}. (Here $J^L_{(-)AR}$ denotes the derivative $(J^L_{(-)A})_{,R}$ etc.). Since
\ber
\amsbb{G}_{AB} = \Omega_{AC}\left[J_{(+)},J_{(-)}\right]^C_{~B}
\eer
the second line may be rewritten as
\ber
\epsilon^+\sigma^{LC}\amsbb{G}_{CA}\nabla^{(-)}_+D_-X^A~,
\eer
where we use the notation $\sigma^{LC}\Omega_{CD}=\delta^C_D$. This part of the variation will cancel the covariant derivative term multiplying 
$\Psi^R_+$ in \re{fullvar}. We are left with
\ber\nn\label{finvarL}
\delta {\cal L} =-\half \epsilon^+D_-\left(\Psi^R_+C_{RR'}\Psi^{R'}_+\right)+ \epsilon^+D_+\left(\Psi^R_+K_{RL}J\Psi^L_-\right)~,
\eer
which ensures invariance of the action. Here where $C_{RR'}$ is the commutator with the canonical complex structure $C_{RR'}:=[J,K_{RR'}]$. 

In deriving \re{finvarL} heavy use is made of integrability and covariant constancy of $J_{(\pm)}$ as well as their explicit expressions \cite{Lindstrom:2005zr}. 

We note that the transformations that leave the action invariant may be written
\ber\nn\label{INV}
&&\delta X^A={\epsilon^+J^A_{(+)B}D_+X^B}+\delta^A_R\epsilon^+\Psi^R_+\\[1mm]\nn
&&\delta \Psi^L_-= \epsilon^+JD_+\Psi^{L}_-
+\epsilon^+J^L_{(-)R'}D_-\Psi^{R'}_+-\epsilon^+J^L_{(-)AR'}\Psi^{R'}_+D_-X^A\\[1mm]\nn
&&~~~~~~~~~-2\epsilon^+K^{LR}\amsbb{G}_{RA}\nabla^{(-)}_+D_-X^A\\[1mm]
&&\delta \Psi^R_+= \epsilon^+J^R_{(+)R'}D_+\Psi^{R'}_+-\epsilon^+J^R_{(+)AR'}\Psi^{R'}_+D_+X^A
\eer
\subsection{$\nff$}
We now investigate if our additional super symmetries \re{CS} can be written in terms of transformations of semichiral fields in  $(2,2)$ superspace.
In our $(1,1)$ language, the relation to semichirals is given by \re{bigpsi} in $(X^L,X^R)$ coordinates\footnote{The reduction of $(\bbX{L},\bbX{R})$.}.  We denote a generic complex structure by $I^A_{~B}$ and write the $X$ transformations as
\ber\nn\label{exinv}
&&\delta X^A=\epsilon^+\left[I^A_{(+)B}D_+X^B+M^A_{~R}\Psi^R_+\right]~,
\eer
where ${\cal L}_1$ is assumed to be invariant (up to total derivatives) under the first transformation on the RHS. Note that $M^A_{~L}=0$.
The formula \re{exinv}  is of the form in \re{INV} and the most general expression compatible with dimensions and symmetries. From \re{bigpsi} we have that
\eject
\ber\nn
&&\delta\Psi^{\dot R}_+=\epsilon^+\left[\left(I^{\dot R}_{(+)R}-[M,J_{(+)}]^{\dot R}_{~R}\right)D_+\Psi^R_+
+\left(I^{\dot R}_{+A,R}+{\cal M}(M,J_{(+)})^{\dot R}_{RA}\right)D_+X^A\Psi^R_+\right.\\[1mm]\nn
&&~~~~~~~~~~~~~~~\left. -M^{\dot R}_{~R,R'}\Psi^{R'}_+\Psi^R_+\right]\\[1mm]\nn
&&\delta\Psi^{\dot L}_-=\epsilon^+\left[-[M,J_{(-)}]^{\dot L}_{~R}D_-\Psi^R_++{\cal M}(M,J_{(-)})^{\dot L}_{RA}D_-X^A\Psi^R_+\right.\\[1mm]\nn
&& ~~~~~~~~~~~~~+(I^{\dot L}_{(+)L}-M^{\dot L}_{~R}J^R_{(+)L})D_+\Psi^L_-+(I^{\dot L}_{(+)A,L}-M^{\dot L}_{~R}J^R_{(+)A,L})D_+X^A\Psi^L_-\\[1mm]\nn
&&~~~~~~~~~~~~~~\left.-M^{\dot L}_{~R,L}\Psi^{L}_-\Psi^R_++\left([I_{(+)},J_{(-)}]^{\dot L}_{~A}-M^{\dot L}_{~R}[J_{(+)},J_{(-)}]^R_{~A}\right)\nabla^{(-)}_+D_-X^A\right]\\[1mm]
&&
\eer
where the Magri-Morosi concomitant for two endomorphisms $I$ and $J$ reads\footnote{Originally defined for a Poisson structure $P$ and a Nijenhuis tensor
$N$ when it reads \cite{MM}
\ber\nn
C^{kj}_m=P^{lj}N^k_{m,l}+P^{kl}N^k_{m,l}-N^l_{m}P^{kj}_{~,l}+N^j_{l}P^{kl}_{~,m}-P^{lj}N^k_{l,m}
\eer and is only a tensor when $[P,N]=0$.}
\ber
{\cal M}(I,J)^{A}_{BD}=I^F_{~B}J^A_{~D,F}-J^F_{~D}I^A_{~B,F}-I^A_{~F}J^F_{~D,B}+J^A_{~F}I^F_{~B,D}~.
\eer
When $M^A_{~R}=\delta^A_R$ and $I_{+}=J_{(+)}$ these transformations reduce to \re{INV}.\\ 

From invariance of \re{completed}, we find a number of relations.
First, raising and lowering indices on $M$ with  $K_{RL}$, 
\ber\nn\label{morecalc}
&&{M_{L[R,\dot R]}}{-}{M_{[R\dot R],L}=0}\\[1mm]\nn
 &&{M_{[R\dot R]}={{-}}\half K_{\dot R \dot L}[I_{(+)},J_{(-)}]^{ \dot L}_{~A}\amsbb{G}^{AL}}K_{LR}\\[1mm]
&&{M^R_{~\dot R}={{-}}\half K_{\dot RL}[I_{(+)},J_{(-)}]^{ L}_{~A}\amsbb{B}^{AR}}
\eer
Note that only the antisymmetric part of $M_{R\dot R}$ is determined by this\footnote{The RHS of the equation containing $M_{[R\dot R]}$ is antisymmetric due to hermiticity conditions.} .
The $D_-$ terms in the variation of $\cal L$ are
\ber\nn
&&\Psi^{\dot R}_+K_{\dot R L}\epsilon^+\left(-[M,J_{(-)}]^L_{~R}D_-\Psi^R_++{\cal M}(M,J_{(-)})^{ L}_{RA}D_-X^A\Psi^R_+\right)
\eer
For this to yield a total $D_-$ derivative, we shall need
\ber\label{dmsym}
&&{-}{K_{(\dot R |L|}[M,J_{(-)}]^L_{~R)}=[J,M]_{(\dot R R)}+C_{(\dot R |R|}M^R_{~R)}=0}
\eer
and 
\ber\label{dmasym}
	&&{K_{[\dot R |L|}{\cal M}(M,J_{(-)})^{ L}_{R]A}D_-X^A=-\half D_-(K_{[\dot R |L|}[M,J_{(-)}]^L_{~R]})}
\eer
Similarily, for the $D_+$ terms to yield a total derivative, we need
\ber\nn\label{Dplus}
&&{K_{LR}I^R_{~\dot R}-K_{\dot R L}I^L_{~L}+K_{LR}[\tilde J,M]^R_{~\dot R}+C_{LL}K^{LR}M_{[R,\dot R]}=0}
\eer
and
\ber\nn\label{finale}
&&{{\left(I^R_{~A,\dot R}+{\cal M}(M,J_{(+)})^R_{\dot RA}\right)K_{RL}+K_{\dot R L B}I^B_{~A}+K_{\dot R L}\left(I^L_{~A,L}-M^L_{~R}J^R_{(+)A,L}\right)}}\\[1mm]\nn
&&{=\left((I^{\dot L}_{~L}-M^{\dot L}_{~R}J_{(+)L}^R)K_{\dot L\dot R}\right)_{,A}}\\[1mm]
\eer
where we have used \re{morecalc} and the explicit form of $J_{(+)}$. 
\section{The $S^3 \times S^1$ model}

\subsection{Duality}
In this section we briefly recapitulate the dualisation of  the BiLP formulation of the $SU(2) \times U(1)$  WZW model \cite{LRRUZ}, albeit in a different version.

We start from the following BiLP potential which gives a sigma model with target space geometry $S^3 \times S^1$;
\ber\label{act1}
K=-ln\hat\chi ln\hat{\bar \chi}+\int^{\frac {\hat\phi\hat{\bar\phi}}{\hat\chi\hat{\bar\chi}}}dq\frac{ln(1+q)}q ~,
\eer
where $\hat \phi$ is chiral, $\bbDB{\pm} \hat \phi=0$, and $\hat \chi$ is twisted chiral, $\bbDB{+} \hat \chi=0=\bbD{-} \hat \chi$.
The potential satisfies the Laplacian
\ber
K_{\hat\phi\hat{\bar\phi}}+K_{\hat\chi\hat{\bar\chi}}=0~,
\eer
and hence the model has $(4,4)$ supersymmetry \cite{Gates:1984nk}. Changing coordinates to new chiral and twisted chiral fields, $\phi=ln\hat\phi, \chi=ln\hat\chi$,
results in
\ber\label{bilp2}
K\to K=-\chi \bar \chi+\int^{\phi+\bar\phi-\chi-\bar\chi}dq~{ln(1+e^q)} ~,
\eer
and makes it  amenable to dualisation of the translation symmetry\footnote{This is equivalent to dualising the scaling symmetry of \re{act1}. These isometries both commute with the extra supersymmetry \cite{Crichigno:2011aa}, \cite{Goteman:2012vp}.}
\ber
\phi\to\phi+\lambda,~\chi \to \chi +\lambda~.
\eer
To apply the gauging prescription of \cite{Lindstrom:2007sq} we add  a term $\alpha(\chi-\bar\chi)(\phi-\bar\phi)$ , which represents a constant $B$-term, to the Lagrangian and rewrite the potential, up to generalized K\"ahler gauge transformations, as
\ber
\half(\chi-\bar\chi)^2+\alpha(\chi-\bar\chi)(\phi-\bar\phi)+\int^{\phi+\bar\phi-\chi-\bar\chi}dq~{ln(1+e^q)}~.
\eer
Following \cite{Lindstrom:2007sq}, we find the first order action (in a WZ gauge);
\ber\label{Vact}
-\half V^2_\chi-\alpha V_\chi V_\phi -V'X'-V_\phi X_\phi-V_\chi X_\chi+\int^{V'}dq~{ln(1+e^q)}~,
\eer
where $V_\phi, V_\chi$ and $V'$ are the Large Vector Multiplet (LVM) fields \cite{Lindstrom:2007sq}\footnote{ In pure gauge they become
\ber\nn
&&V_\phi=i(\bar \phi-\phi)\\[1mm]\nn
&&V_\chi=i(\bar \chi-\chi)\\[1mm]\nn
&&V'=\phi+\bar\phi-\chi-\bar\chi~.
\eer},
and the Lagrange multipliers are combinations of semichiral fields
\ber\nn
&&X_\phi=\ihalf(\ell-\bar\ell-r+\bar r)\\[1mm]\nn
&&X_\chi=\ihalf(-\ell+\bar\ell-r+\bar r)\\[1mm]
&&X'=\half(\ell+\bar\ell-r-\bar r)~.
\eer
Eliminating the LVM and massaging the integral we find the dual semichiral action in the form
\ber\label{mod1}
-\frac 1{2\alpha^2}X_\phi^2+\frac 1{\alpha}X_\phi X_\chi -\int ^{X'}dq~ ln(e^q-1)~.
\eer
This is the potential that is expected to have additional supersymmetries due to those of the dual BiLP model.

\subsection{The geometry}
The reduction of a semichiral model to $(1,1)$ superspace may be expressed in several useful coordinate systems. E.g., the $(X^L,X^R)$ coordinates directly obtained in the reduction is related to the $(X,Y)$ system where $J_{(+)}$ is canonical via a coordinate transformation \cite{Bogaerts:1999jc},\cite{Lindstrom:2005zr}.

We now derive the metric in $(X,Y)$ coordinates for \re{mod1} . To this end we first calculate the various ingredient matrices according to the formulae in \cite{Lindstrom:2005zr}. 

Without loss of generality, we set $\alpha$ to $-1$ in  \re{mod1} and drop the prime on $X'$. The $Y$ coordinates are defined to be $K_L=:Y$.
We find 
\ber\nn\label{Ytfs}
&&-y=K_\ell=\ihalf X_\chi +\half ln (e^{X}-1)=\quart (\ell-\bar\ell)+\quart(r-\bar r)+\half  ln (e^X-1)\\[1mm]\nn
&&\Rightarrow\\[1mm]\nn
&& X=ln(1+e^{-(y+\bar y)})\\[1mm]\nn
&&r+\bar r=\ell+\bar \ell-2ln(1+e^{-(y+\bar y)})~,\\[1mm]
&&r-\bar r=-2(y-\bar y)-(\ell-\bar\ell)
\eer
The relevant matrices of derivatives of $K$ are
\ber\nn\label{HESS}
&&K_{LL}=-\frac 1 {4N} \left(\begin{array}{cc}E&1\\1&E\end{array}\right)~,~~~~-K_{RR}=\frac 1 {4N} \left(\begin{array}{cc}2-E&M\\4N+1&2-E\end{array}\right)\\[2mm]\nn
&&-K_{LR}=-\frac 1 {4N} \left(\begin{array}{cc}1&E\\E&1\end{array}\right)~,~~~~-K^{RL}=\frac 1 {e^X} \left(\begin{array}{cc}1&-E\\-E&1\end{array}\right)\\[2mm]
&&-C_{LL}=\frac {2i} {4N} \left(\begin{array}{cc}0&1\\-1&0\end{array}\right)~,~~~~-C_{RR}=\frac {2iM} {4N} \left(\begin{array}{cc}0&1\\-1&0\end{array}\right)
\eer
where we have introduced the notation
\ber\label{DEFs}
N:=e^{-(y+\bar y)}=e^X-1~,~~E:=2e^X-1~,~~M:=4N+1~,
\eer
for combinations that will occur frequently in our formulae.
The  metric and $B$-field in $(X,Y)$ coordinates can be calculated from the formulae in \cite{Lindstrom:2005zr}:
\ber\nn
&&-\amsbb{E}_{LL}=J(K_{LL}K^{LR}JK_{RL}-K_{LR}JK^{RL}K_{LL}-K_{LL}K^{LR}C_{RR}K^{RL}K_{LL})=-2\sigma_1\\[1mm]\nn
&&\amsbb{E}_{LY}=J(K_{LR}JK^{RL}+K_{LL}K^{LR}C_{RR}K^{RL})=-\frac 1{e^X} 
\left(\begin{array}{cc}2-e^X&-2E\\
-2E&2-e^X\end{array}\right)\\[1mm]\nn
&&\amsbb{E}_{YL}=J(-K^{LR}JK_{RL}+K^{LR}C_{RR}K^{RL}K_{LL})=\frac 1{e^X} 
\left(\begin{array}{cc}2-e^X&2E\\
2E&2-e^X\end{array}\right)\\[1mm]\nn
&&-\amsbb{E}^{YY}=-JK^{LR}C_{RR}K^{RL}=-\frac{2M}{e^X}\sigma_1~.
\eer
Note that the tensor $\amsbb{E}$ depends on $y+\bar y$ only.
From the formulae for $\amsbb{E}=\amsbb{G}+\amsbb{B}$, it follows that the metric is\footnote{Since a lot of the objects have $2D$ complex submatrices, it is convenient to introduce the Pauli matrices $\sigma_i$ and write matrices as direct products.}
\ber\label{semetric}
\amsbb{G}=\frac 2{e^X}\left(\begin{array}{cc}e^{-X}& E\\
 E&M\end{array}\right)\otimes \sigma_1~,
\eer
with inverse
\ber\label{semetricinv}\nn
&&\amsbb{G}^{-1}=-\frac {e^X}{2N}\left(\begin{array}{cc}-M&E\\
 E&-{e^X}\end{array}\right)\otimes \sigma_1
 =:-\frac {e^X}{2N}{\tt h}\otimes \sigma_1~.
\eer
Also
\ber
\amsbb{B}_{YL}=(2e^{-X}-1)\mathbb{1}~.
\eer
In these coordinates $J_{(+)}$ is canonical while
\ber\nn\label{jminus}
J_{(-)}&=e^{-X}\left(
\begin{array}{cc}
({2}-e^X)J-2E\sigma_2&-2M\sigma_2\\
  2e^X\sigma_2&({2}-e^X)J+2E\sigma_2 \end{array}
 \right)\\[2mm]
 &=:e^{-X}\left(2{\tt j}\otimes \sigma_2+(2-e^X)\otimes i\sigma_3\right)~.
\eer
where we use that $J=i\sigma_3$.

\subsection{Additional SUSY in $(1,1)$}
\subsubsection{Deriving the transformations on the $(1,1)$ coordinate fields}

In the original BiLP, \re{act1} the additional super symmetries read \cite{Gates:1984nk}
\ber\nn
&&\delta \hat \phi=\bar\epsilon^+\bbDB{+}\hat{\bar\chi}+\bar\epsilon^-\bbDB{-}\hat\chi\\[1mm]\nn
&&\delta \hat {\bar\phi}=\epsilon^+\bbD{+}\hat{\chi}+\epsilon^-\bbD{-}\hat{\bar\chi}\\[1mm]\nn
&&\delta \hat {\chi}=-\bar\epsilon^+\bbDB{+}\hat{\bar\phi}-\epsilon^-\bbD{-}\hat{\phi}\\[1mm]
&&\delta \hat {\bar\chi}=-\epsilon^+\bbD{+}\hat{\phi}-\bar\epsilon^-\bbDB{-}\hat{\bar\phi}~,
\eer
and in the transformed version \re{bilp2} they become
\ber\nn\label{2ndsusy}
&&\delta \phi=e^{\bar\chi-\phi}\bar\epsilon^+\bbDB{+}{\bar\chi}+e^{\chi-\phi}\bar\epsilon^-\bbDB{-}\chi\\[1mm]\nn
&&\delta {\bar\phi}=e^{\chi-\bar\phi}\epsilon^+\bbD{+}{\chi}+e^{\bar\chi-\bar\phi}\epsilon^-\bbD{-}{\bar\chi}\\[1mm]\nn
&&\delta {\chi}=-e^{\bar\phi-\chi}\bar\epsilon^+\bbDB{+}{\bar\phi}-e^{\phi-\chi}\epsilon^-\bbD{-}{\phi}\\[1mm]
&&\delta {\bar\chi}=-e^{\phi-\bar\chi}\epsilon^+\bbD{+}{\phi}-e^{\bar\phi-\bar\chi}\bar\epsilon^-\bbDB{-}{\bar\phi}~.
\eer
These relations survive in the $(1,1)$ reduction with $\bbD{\pm}\to D_\pm$.
From \re{2ndsusy} we then read off the additional complex structures according to

\ber
\delta \varphi=\half\left[ \left (J^{(1)}_{(\pm)}+iJ^{(2)}_{(\pm)}\right)\epsilon^\pm D_\pm\varphi+\left (J^{(1)}_{(\pm)}-iJ^{(2)}_{(\pm)}\right)\bar\epsilon^\pm D_\pm\varphi\right]
\eer\nn
For the $J^{(\textfrak a)}_{(+)}$ we find
\ber
&&J^{(1)}_{(+)}=\left(\begin{array}{cccc}
0&0&0&e^{\bar\chi-\phi}\\
0&0&e^{\chi-\bar\phi}&0\\
0&-e^{\bar\phi-\chi}&0&0\\
-e^{\phi-\bar\chi}&0&0&0\end{array}\right)\\[2mm]
&&J^{(2)}_{(+)}=\left(\begin{array}{cccc}
0&0&0&ie^{\bar\chi-\phi}\\
0&0&-ie^{\chi-\bar\phi}&0\\
0&-ie^{\bar\phi-\chi}&0&0\\
ie^{\phi-\bar\chi}&0&0&0\end{array}\right)~,
\eer
with $J^{(3)}_{(+)}=J$, the canonical complex structure.

We would like to see what these complex structures look like in $(1,1)$ coordinates related to the semichiral description.  While T-dual formulations are not in general related by coordinate transformations, they are in this case due to the special choice of the isometry direction; we have dualised along the common $U(1)$. We now need to find the coordinate transformation. To this end, we note that the relations \re{ctfs}, derived in the appendix,
\ber\nn
&&V_\phi=X_\chi+X_\phi\\[1mm]\nn
&&V_\chi=X_\phi\\[1mm]
&&V=ln(e^X-1)~.
\eer
do not completely determine the transformations. Identifying the LHS with BiLP fields (WZ-gauge) and writing out the RHS we have
\ber\nn\label{ctfs}
&&i(\bar \phi-\phi)=-i (r-\bar r)\\[1mm]\nn
&&i(\bar\chi-\chi)=\ihalf(\ell-\bar\ell-r+\bar r)\\[1mm]
&&\phi+\bar\phi-\chi-\bar\chi=ln(e^{\half(\ell+\bar \ell-r-\bar r)}-1)~.
\eer
It turns out to be most convenient to identify the coordinate transformation to $(X,Y)$ coordinates where the complex structure derived from the semi side, $J_{(+)}$, is canonical\footnote{The map of $J_{(\pm)}$ under duality is discussed in \cite{Ivanov:1994ec} where the dual model appears in some preferred coordinates.  We have not investigated  the relation to the present coordinates.}.  A coordinate transformation to $(X^L,X^R)$ coordinates will then give a non-canonical $J_{(+)}$. Comparing this to \re{Ytfs}, we identify
\ber\nn
&&y=\chi-\phi\\[1mm]
&&\ell-\bar\ell=2(\bar\chi-\chi)-(\bar \phi-\phi)~,
\eer
but $\ell+\bar \ell$ is left undetermined.  In both coordinate systems we have $J^{(3)}_{(+)}=J$. Requiring that the coordinate transformation takes the canonical complex structure into itself determines $\ell+\bar \ell=\phi +\bar \phi-2(\bar\chi-\chi)$, which results in
\ber
\ell=\phi-2\chi~.
\eer
This gives the transformation Jacobian 
\ber
&&\mathbb{\Lambda} =\left(\begin{array}{cc}
\frac{\partial L}{\partial \phi}&\frac{\partial L}{\partial \chi}\\
\frac{\partial Y}{\partial \phi}&\frac{\partial Y}{\partial \chi}\end{array}\right)=\left(\begin{array}{cc}
\mathbb{1}&-2\mathbb{1}\\
-\mathbb{1}&\mathbb{1}\end{array}\right)~,
\eer
with inverse
\ber
\mathbb{\Lambda}^{-1} =-\left(
\begin{array}{cc}
\mathbb{1}&2\mathbb{1}\\
\mathbb{1}&\mathbb{1}
\end{array}\right)~,
\eer
These transformations correctly relates the BiLP metric derived from \re{bilp2} to  the semi metric \re{semetric}.
We now write the extra complex structures $J^{(\textfrak a)}_{(+)}$ as 
\ber
&&J^{(\textfrak a))}_{(+)}=\left(\begin{array}{cc}
0&\amsbb{A}^{(\textfrak a)}\\
-(\amsbb{A}^{(\textfrak a)})^{-1}&0\end{array}\right)
\eer
for $\textfrak a=1,2$, with
\ber\nn\label{Adef}
&&\amsbb{A}^{(1)}=\left(\begin{array}{cc}
0&e^{\bar\chi-\phi}\\
e^{\chi-\bar\phi}&0\end{array}\right)=\left(\begin{array}{cc}
0&e^{\varphi + y}\\
e^{-\varphi+\bar y}&0\end{array}\right)\\[2mm]
&&\amsbb{A}^{(2)}=\left(\begin{array}{cc}
0&ie^{\bar\chi-\phi}\\
-ie^{\chi-\bar\phi}&0\end{array}\right)=\left(\begin{array}{cc}
0&ie^{\varphi + y}\\
-ie^{-\varphi+\bar y}&0\end{array}\right)~,
\eer
where
\ber
\varphi:=(\ell-\bar \ell)+(y-\bar y)~.
\eer
The expressions for $J^{(\textfrak a)}_{(+)}$ in $(X,Y)$ coordinates then become
\ber\nn\label{CS}
&&J^{(\textfrak a)}_{(+)}=
\left(\begin{array}{cc}
-E&-M\\
e^X&E\end{array}\right)\otimes\amsbb{A}^{(\textfrak a)}=:{\tt j}\otimes\amsbb{A}^{(\textfrak a)}~,\\[2mm]
&&~~~~{\tt j}^2=-N~, ~~~~~~(\amsbb{A}^{(\textfrak a)})^2=N^{-1}~,
\eer
where we again use the notation in \re{DEFs}. The complex structures $J^{(\textfrak a)}_{(+)}$ preserve the metric \re{semetric}, as confirmed by an explicit calculation. %To this end it is useful to note that the function {\tt g} in

\subsubsection{$(X^L,X^R)$ coordinates}

Using \re{HESS} in the Jacobian
\ber\nn
&&\mathbb{\Lambda}=\left(\begin{array}{cc}
\mathbb{1}&0\\
K_{LL}&K_{LR}\end{array}\right)
=\frac 1 {4N} \left(\begin{array}{cc}
4N&0\\
-(E\mathbb{1}+\sigma_1)&\mathbb{1} +E\sigma_1\end{array}\right)~,\\[2mm]
&&\mathbb{\Lambda}^{-1}=\left(\begin{array}{cc}
\mathbb{1}&0\\
-K^{RL}K_{LL}&K^{RL}\end{array}\right)=\left(\begin{array}{cc}
\mathbb{1}&0\\
\sigma_1&-e^{-X}(\mathbb{1} - E\sigma_1)\end{array}\right)~,
\eer
we find the expressions for $J^{(\textfrak a)}_{(+)}$ in left right coordinates $(X^{L},X^{R})$: 
\ber\nn
J^{(\textfrak a)}_{(+)}=&&\\[1mm]\nn
&&\frac 1 {4N}\left\{\left(\begin{array}{cc}E&-M\\
e^{-X}&-e^{-X}E
\end{array}\right)\otimes {\mathbb{A}}^{(A)}+\left(\begin{array}{cc}M&-ME\\
e^{-X}E&-e^{-X}E^2
\end{array}\right)\otimes {\mathbb{A}}^{(A)}\sigma_1\right.\\[2mm]
&&~~~~~~\left. +\frac N{e^X}\left[
\left(\begin{array}{cc}0&0\\
1&-E
\end{array}\right)\otimes \bar{\mathbb{A}}^{(A)}+\left(\begin{array}{cc}0&0\\
E&-1
\end{array}\right)\otimes \bar{\mathbb{A}}^{(A)}\sigma_1
\right]\right\}~,
\eer
\normalsize where $Y(X^{L},X^{R})$ is given by the relations in \re{Ytfs}.

\section{Examples}
\subsection{Hyperk\"ahler}
We want to show that there are hyperk\"ahler solutions to our problem in $4D$. 
To this end we note that, when $c$ is constant
we have available the following hyperk\"ahler structure \cite{Goteman:2009xb};
\ber
{\cal I}:= J_{(+)}~, ~~{\cal J}:= \frac 1 {\sqrt{1-c^2}}(J_{(-)}+cJ_{(+)})~,~~{\cal K}:=\frac 1 {2\sqrt{1-c^2}}[J_{(+)},J_{(-)}]~.
\eer
The relations \re{morecalc} determine $M$ in the three cases according to
\ber\nn\label{HKE}
{\cal I}: && M^R_{~\dot R} = \delta^R_{~\dot R}~,~~M_{[\dot R R]}=0\\[1mm]\nn
{\cal J}: && M^R_{~\dot R} = \frac {c~\delta^R_{~\dot R}}{\sqrt {1-c^2}}~,~~\\[1mm]\nn
{\cal K}: && M^R_{~\dot R}K_{L R} = -\frac 1 {\sqrt{1-c^2}} K_{\dot R L}J_{(-)L}^L=-\frac 1 {\sqrt{1-c^2}}JK_{\dot R L}\\[1mm]\nn
&& M_{[\dot R R]} =-\frac 1 {\sqrt{1-c^2}} K_{\dot R L}J_{(-)R}^L= -\frac 1 {\sqrt{1-c^2}} C_{\dot R R}\\[1mm]
\eer
Each case satisfies the first relation in \re{morecalc} (provided that $c$ is constant) .\\
The conditions \re{dmsym} is satisfied by the hyperk\"ahler structure \re{HKE}.
The relation  \re{dmasym} is satisfied for ${\cal I}$ and ${\cal J}$  by direct insertion. 
For ${\cal K}$ we determine the full 
\ber M^L_{~\dot R}= -\frac 1 {\sqrt{1-c^2}}K^{LR}JK_{R\dot R}
\eer
and find that 
\ber
K_{[\dot R |L|}[M,J_{(-)}]^L_{~R]}=0~,
\eer
and the issue becomes the vanishing of $\cal{M}$.  This is again confirmed by direct insertion of the ${\cal K}$ expressions from \re{HKE}.

As a final check we also find that
the relations \re{Dplus} and \re{finale} are indeed satisfied for  ${\cal I}, {\cal J}$ and ${\cal K}$.

\subsection{$SU(2)\otimes U(1)$}

Using \re{CS} and \re{jminus} we find that in $(X,Y)$ coordinates\\
\ber
[J^{(\textfrak a)}_+,J_-]=-2e^{-X}\left((2-e^X){\tt j}\otimes(b^{(\textfrak a)}\sigma_1+a^{(\textfrak a)}\sigma_2)+2N\mathbb{1}\otimes a^{(\textfrak a)}i\sigma_3\right)~,
\eer
where {\tt j} is defined in \re{jminus}. Using \re{Adef} we have defined
\ber\nn\label{psistuff}
&&\amsbb{A}^{(1)}=:a^{(1)}\sigma_1-b^{(1)}\sigma_2=:\frac 1 {\sqrt{N}}(cos\psi ~\!\sigma_1-sin\psi~\! \sigma_2)\\[1mm]
&&\amsbb{A}^{(2)}=:a^{(2)}\sigma_1-b^{(2)}\sigma_2=:\frac 1 {\sqrt{N}}(-sin\psi ~\!\sigma_1-cos\psi~\! \sigma_2)~,
\eer
and
\ber\label{psidef}
i\psi:=(\ell-\bar \ell)+\frac  32(y-\bar y)~.
\eer
As is clear from \re{morecalc}, we shall need $[J^{(\textfrak a)}_+,J_-]{\amsbb{G}}^{-1}$. We find
\ber
[J^{(\textfrak a)}_+,J_-]{\amsbb{G}}^{-1}=
2a^{(A)}{\tt h}
\otimes\sigma_2+(2-e^X)
\left(\begin{array}{cc}0&1\\-1&0\end{array}\right)
\otimes\left(b^{(\textfrak a)}\mathbb{1}-a^{(\textfrak a)}i\sigma_3\right)~,
\eer
where {\tt h} is defined in \re{semetricinv}.
In $(X^L,X^R)$ coordinates this becomes
\ber\nn
(2e^{-X}-1)\left(\begin{array}{cc}0&1\\-1&0\end{array}\right)\otimes b^{(\textfrak a)}(\mathbb{1}-E\sigma_1)+\left(\begin{array}{cc}2M&E\\E&0\end{array}\right)\otimes a^{(\textfrak a)}\sigma_2-\left(\begin{array}{cc}0&1\\-1&0\end{array}\right)\otimes a^{(\textfrak a)}i\sigma_3\\[2mm]~.
&&
\eer
We read off the matrices relevant to \re{morecalc}
\ber\nn\label{wzwm}
&&M_{[R\dot R]}=\half K_{\dot R \dot L}\left([J^{(\textfrak a)}_+,J_-]{\amsbb{G}}^{-1}\right)^{\dot L L}K_{LR}=-\frac{Me^Xa^{(A)}}{4N}\sigma_2\\[1mm]\nn
&&M_{LR}=\half K_{R \dot L}\left([J^{(\textfrak a)}_+,J_-]{\amsbb{G}}^{-1}\right)^{\dot L R}K_{RL}\\[1mm]
&&=-\half\frac {e^X}{4N}\left[(2e^{-X}-1)b^{(\textfrak a)}(1+E\sigma_1)+a^{(\textfrak a)}(E\sigma_2-i\sigma_3)\right]
\eer
We find that  the quantities in \re{wzwm} indeed satisfy the first relation in \re{morecalc}. Proceeding to \re{dmsym} and \re{dmasym}, we find that \re{dmsym} is also satisfied using \re{wzwm}, and that the $b^{(\textfrak a)}$ terms in \re{dmasym}  cancel. However, the remaining terms in \re{dmasym} must satisfy
\ber\nn
&&\left(M^F_{~[r}K_{\bar r ]F}\right)_R=0\\[1mm]
&&M^r_{~r[,r}K_{\bar r ]L}=0~,
\eer
where knowledge of the form of $J_{(-)}$ along with partial information from \re{wzwm} has been used.
While the first of these equations determines the remaining parts of $M^L_{~R}$, the second equations must be identically satisfied by $M^R_{~R}$ in \re{wzwm}. This is not the case.
\section{Discussion}

We have extended the  $(1,1)$ formulation of semichiral sigma models to allow for a treatment of extra super symmetries with on-shell closure. To exemplify the general method we have shown that a set of hyperk\"ahler geometries arise as solutions of the conditions for extra supersymmetry.  We have further constructed the extra super symmetries in a  semichiral models dual to a BiLP model with ``manifest'' $(4,4)$ susy on the BiLP side. 
This model fails the criteria for the additional supersymmetry to be manifest as transformations of $(2,2)$ semichirals. Another way of saying this is that the $(4,4)$ supersymmetry is incompatible with the introduction of the $(2,2)$ auxiliary spinor fields. The key ingredient in the analysis is to show that invariance of the action fails (on-shell closure of the algebra is ensured by construction). Note that the analysis shows that not even an extra supersymmetry of one handedness only is possible.

 Our analysis is carried out at the $(1,1)$ level, where conditions for additional supersymmetries are well established since  thirty years \cite{Gates:1984nk}. 

An analysis at the $(2,2)$ level already indicated that the remedy suggested in \cite{Goteman:2012qk} will not work; a formulation including central charge transformations will typically display the original obstructions when we go on-shell.

A further indication of problems with an extra supersymmetry comes from dualisation procedure itself. One would expect the parent action, where the chirality constraints on the chiral and twisted chiral superfields  have been relaxed, to have the extra supersymmetry. This would mean the the LVM gauge multiplet could carry extra supersymmetry. This was concluded to be impossible under fairly general assumptions in \cite{Goteman:2010sf}.

In view of this result, it is reasonable to conjecture  that manifest extra supersymmetries involving semichiral fields together with a $4D$ target space is only possible in  models including auxiliary fields such as in the  $(4,4)$  superspace setting of \cite{Lindstrom:1994mw}.
\begin{flushleft}
\bigskip

Acknowledgement: Discussions with M. Ro\v cek at various stages of this work are gratefully acknowledged.
Supported in part by VR grant 621-2013-4245
\end{flushleft}

\appendix
\section{Duality in $(1,1)$}

In this section we reduce the action \re{Vact} (with $\alpha = -1$)  to $(1,1)$ and eliminate the LVM there instead. This makes clear the issue of coordinate transformations at the $(1,1)$ level. We replace covariant derivatives according to
\ber
\bar D_\pm\to D_\pm+iQ_\pm~,~~ D_\pm\to D_\pm-iQ_\pm~.
\eer
To facilitate the calculation we introduce the following notation:
\ber\nn
&&Y_\pm^A:=Q_\pm X_A~,\\[1mm]\nn
&&Z^A:=Q_+Q_\pm X_A~,~~A=\phi,~\chi,~X\\[1mm]\nn
&&s^\ell:=\ell+\bar\ell~,~~d^\ell:=\ell-\bar\ell\\[1mm]\nn
&&s^r:=r+\bar r~,~~d^r:=r-\bar r\\[1mm]\nn
&&\Sigma:=\psi+\bar\psi\\[1mm]
&&\Lambda:=\psi-\bar\psi
\eer
and define the $(1,1)$ components of the LVM (in WZ gauge) as
\ber\nn
&&V_\chi  |=:V_\chi ~,~~Q_\pm V_\chi  |=:(A+B)_\pm~,~~Q_+Q_\pm V_\chi  |=:F\\[1mm]\nn
&&V_\phi  |=:V_\phi ~,~~Q_\pm V_\phi  |=:B_\pm~,~~Q_+Q_\pm V_\phi  |=:G\\[1mm]
&&V_X  |=:V_X~,~~Q_\pm V_X |=:C_\pm~,~~Q_+Q_\pm V_X  |=:H~.
\eer
The Lagrangian becomes
\ber\nn
&&F(V_\phi-V_\chi-X_\chi)+G(V_\chi-X_\phi)+H(ln (1+e^V)-X)\\[1mm]\nn
&&-V_\phi Z^\phi-V_\chi Z^\chi-V'Z^X-(A_++Y^\chi_+)(A_-+Y_-^\chi)+Y^\chi_+Y_-^\chi\\[1mm]\nn
&&+(B_++Y^\phi_++Y^\chi_+)(B_-+Y^\phi_-+Y^\chi_-)-(Y^\phi_++Y^\chi_+)(Y^\phi_-+Y^\chi_-)\\[1mm]\nn
&&+\left(C_+-Y^\chi_+\left(\frac {1+e^V}{e^V}\right)\right)\left(\frac {e^V}{1+e^V}\right)\left(C_--\left(\frac {1+e^V}{e^V}\right)Y^\chi_-\right)\\[1mm]
&&-Y^X_+Y^X_-\left(\frac {1+e^V}{e^V}\right)~.
\eer
Integrating out $F,G,H$ gives the coordinate transformation
\ber\nn\label{ctfs}
&&V_\phi=X_\chi+X_\phi\\[1mm]\nn
&&V_\chi=X_\phi\\[1mm]
&&V=ln(e^X-1)~.
\eer
Integrating $V_\phi=, V_\chi, V$ determines $F, G, H$ in terms of the components of the semis:
\ber\nn
&&F=Z^\phi\\[1mm]\nn
&&G=Z^\phi+Z^\chi \\[1mm]
&&H=\left(\frac {1+e^V}{e^V}\right)\left[ \left(1-\frac {1+e^V}{e^V}\right)Y^X_+Y^X_-+Z^X\right]~.
\eer
(No contribution from $C_\pm$ terms et.c.. 1.5 formalism). Finally, integrating $A,B,C$ again determines these fields in terms of the semi components. This leaves us with a purely semi Lagrangian;
\ber\nn
&&Y^\chi_+Y_-^\chi-(Y^\phi_++Y^\chi_+)(Y^\phi_-+Y^\chi_-)-Y^X_+Y^X_-\left(\frac {e^X}{e^X-1}\right)\\[1mm]
&&-(X_\chi+X_\phi )Z^\phi-X_\phi Z^\chi-ln(e^X-1)Z^X
\eer
As a check that this agrees with the reduced semi action, we integrate out the auxiliary spinors $\Psi_\pm$ and reconstruct the complex structures $J^{(\pm)}$. We shall need
\ber\nn
&&Y^\phi_+=\ihalf\left[iD_+s^\ell-\Lambda^r_+\right]\\[1mm]\nn
&&Y^\phi_-=\ihalf\left[\Lambda^\ell_--iD_-s^r\right]\\[1mm]\nn
&&Y^\chi_+=-\ihalf\left[iD_+s^\ell+\Lambda^r_+\right]\\[1mm]\nn
&&Y^\chi_-=-\ihalf\left[\Lambda^\ell_-+iD_-s^r\right]\\[1mm]\nn
&&Y^X_+=\half\left[iD_+d^\ell-\Sigma^r_+\right]\\[1mm]\nn
&&Y^X_-=\half\left[\Sigma^\ell_--iD_-d^r\right]\\[1mm]\nn
&&Z^\phi=-\half\left[D_+\Sigma^\ell_-+D_-\Sigma^r_+\right]\\[1mm]\nn
&&Z^\chi=\half\left[D_+\Sigma^\ell_--D_-\Sigma^r_+\right]\\[1mm]
&&Z^X=\ihalf\left[D_+\Lambda^\ell_-+D_-\Lambda^r_+\right]
\eer
From the variations we find:
\ber\nn
&&\delta \Lambda^r_+:~~~~~\Lambda^\ell_-=3iD_-s^r-2i\left(\frac {e^X}{e^X-1}\right)D_-X\\[1mm]\nn
&&\delta \Sigma^r_+:~~~~~\Sigma^\ell_-=iD_-d^r-2\left(\frac {e^X-1}{e^X}\right)D_-(X_\chi+2X_\phi)
\\[1mm]\nn
&&\delta \Lambda^\ell_-:~~~~~\Lambda^r_+=-iD_+s^\ell+2i\left(\frac {e^X}{e^X-1}\right)D_+X\\[1mm]\nn
&&\delta \Sigma^\ell_-:~~~~~\Sigma^r_+=iD_+d^\ell+2\left(\frac {e^X-1}{e^X}\right)D_+X_\chi~.
\\[1mm]\nn
\eer
This implies that (in the notation of \re{DEFs})
\ber\nn
&&\Psi^\ell_-=i\frac 1 {4e^XN}\left[(1+E^2)D_-\ell+2ED_-\bar \ell-2(4N+1)ED_-r-2(4N+1)D_-\bar r\right]]\\[1mm]
&&\Psi^r_+=i\frac 1 {4e^XN}\left[2ED_+\ell+2D_+\bar\ell-(1+E^2)D_+r-2ED_+\bar r\right]
\eer
These are the correct expressions for these auxiliary fermions, as may be checked using the matrices \re{HESS} in the formulae for $J^{(\pm)}$ in \cite{Lindstrom:2005zr}.

\section{An alternative dual form}

We have been studying the action \re{mod1}. It can be cast into a different form which connects to the results in \cite{Sevrin:2011mc}. We perform a  Legendre transformation of the right semichiral 
superfields (for $\alpha=1$)
\ber
\tilde K=K(L, x,\bar x)-xr-\bar x\bar r
\eer
which together with the change $L\to-L$ (and some manipulations of the integral)  brings the potential to the form found in \cite{Sevrin:2011mc}:
\ber\label{mod2}
-(\ell-\bar r)(\bar \ell-r) +\int^{r+\bar r}dq~ln(1+e^q)~.
\eer
We already know the metric in these $(X^L,X^R)$ coordinates from \cite{Sevrin:2011mc}
\ber
\amsbb{G}=\left(\begin{array}{cc}\sigma_1&-Z\\
-Z&Z\sigma_1\end{array}\right)~,
\eer
where 
\ber\label{Z}
Z:=\frac 1 {1+e^{r+\bar r}}=\frac 1 {1+e^{\ell+\bar \ell-y-\bar y}}~,
\eer
and transformation to new $(X,Y)$ coordinates  reads
\ber\nn
&&y=K_\ell=\bar\ell-r\\[1mm]
&&r=\bar\ell-y~.
\eer
The corresponding Jacobian is \cite{Lindstrom:2005zr} 
\ber
&&J=\left(\begin{array}{cc}\mathbb{1}&0\\
-K^{RL}K_{LL}&K^{RL}\end{array}\right)~.
\eer
We have
\ber\nn
&&K_{LL}=-\sigma_1~,~~~~K_{RR}=-Z\sigma_1\\[1mm]\nn
&&K_{LR}=\mathbb{1}~,~~~~K^{RL}=\mathbb{1}\\[1mm]~'
\eer
which implies
\ber\label{met2}
\amsbb{G}\to\left(\begin{array}{cc}-\sigma_1&0\\
0&Z\sigma_1\end{array}\right)
\eer


\begin{thebibliography}{6666}
 
  %\cite{Lindstrom:2005zr}
\bibitem{Lindstrom:2005zr} 
  U.~Lindstrom, M.~Rocek, R.~von Unge and M.~Zabzine,
  ``Generalized Kahler manifolds and off-shell supersymmetry,''
  Commun.\ Math.\ Phys.\  {\bf 269}, 833 (2007)
  [hep-th/0512164].
  %%CITATION = HEP-TH/0512164;%%
  
     %\cite{Goteman:2010sf}
\bibitem{Goteman:2010sf} 
  M.~Goteman, U.~Lindstrom, M.~Rocek and I.~Ryb,
 ``Off-shell N=(4,4) supersymmetry for new (2,2) vector multiplets,''
  JHEP {\bf 1103}, 088 (2011)
  [arXiv:1008.3186 [hep-th]].
  %%CITATION = ARXIV:1008.3186;%%

  
  %\cite{Goteman:2012qk}
\bibitem{Goteman:2012qk} 
  M.~Goteman, U.~Lindstrom and M.~Rocek,
``Semichiral Sigma Models with 4D Hyperkaehler Geometry,''
  JHEP {\bf 1301}, 073 (2013)
  [arXiv:1207.4753 [hep-th]].
  %%CITATION = ARXIV:1207.4753;%%
  
  %\cite{Rocek:1991vk}
\bibitem{Rocek:1991vk} 
  M.~Rocek, K.~Schoutens and A.~Sevrin,
  ``Off-shell WZW models in extended superspace,''
  Phys.\ Lett.\ B {\bf 265}, 303 (1991).
  %%CITATION = PHLTA,B265,303;%%
  
  \bibitem{LRRUZ}
  U.~Lindstr\"om, I.~Ryb, M.~Ro\v cek, R.von Unge, M.~Zabzine,
  ``T-duality for the $S^1$ piece in the $S^3 \times S^1$ model,''
  October 2009, unpublished.
  
     %\cite{Bakas:1995hc}
\bibitem{Bakas:1995hc} 
  I.~Bakas and K.~Sfetsos,
  ``T duality and world sheet supersymmetry,''
  Phys.\ Lett.\ B {\bf 349}, 448 (1995)
  [hep-th/9502065].
  %%CITATION = HEP-TH/9502065;%%


   %%\cite{Gates:1984nk}
\bibitem{Buscher:1987uw} 
  T.~Buscher, U.~Lindstr\"om and M.~Ro\v cek,
  ``New supersymmetric sigma models with Wess Zumino terms,''
  Phys.\ Lett.\ B {\bf 202}, 94 (1988).
  %%CITATION = PHLTA,B202,94;%%
  
    %\cite{Gates:1984nk}
\bibitem{Gates:1984nk} 
  S.~J.~Gates, Jr., C.~M.~Hull and M.~Rocek,
``Twisted Multiplets and New Supersymmetric Nonlinear Sigma Models,''
  Nucl.\ Phys.\ B {\bf 248}, 157 (1984).
  %%CITATION = NUPHA,B248,157;%%

 %\cite{Gualtieri:2003dx}
\bibitem{Gualtieri:2003dx}
  M.~Gualtieri,
  ``Generalized complex geometry,''
  Oxford University DPhil thesis,
  [arXiv:math/0401221].
  %%CITATION = MATH/0401221;%%
  
 \bibitem{MM}
 F.~Magri  and C.~Morosi, 
 ``A geometrical characterization of integrable Hamiltonian systems through the theory of Poisson-Nijenhuis manifolds'',
 Quadrini del Dipartimento di Matematica, Universita di Milano, 1984.
 
   
   %\cite{Crichigno:2011aa}
\bibitem{Crichigno:2011aa} 
  P.~M.~Crichigno,
  ``The Semi-Chiral Quotient, Hyperkahler Manifolds and T-Duality,''
  JHEP {\bf 1210}, 046 (2012)
  [arXiv:1112.1952 [hep-th]].
  %%CITATION = ARXIV:1112.1952;%%
  
    %\cite{Goteman:2012vp}
\bibitem{Goteman:2012vp} 
  M.~Goteman,
``N=(4,4) supersymmetry and T-duality,''
  Symmetry {\bf 4}, 603 (2012)
  [arXiv:1208.2166 [hep-th]].
  %%CITATION = ARXIV:1208.2166;%%
  
   %\cite{Lindstrom:2007sq}
\bibitem{Lindstrom:2007sq} 
  U.~Lindstrom, M.~Rocek, I.~Ryb, R.~von Unge and M.~Zabzine,
  ``T-duality and Generalized Kahler Geometry,''
  JHEP {\bf 0802}, 056 (2008)
  [arXiv:0707.1696 [hep-th]].
  %%CITATION = ARXIV:0707.1696;%%
  
%\cite{Bogaerts:1999jc}
\bibitem{Bogaerts:1999jc} 
  J.~Bogaerts, A.~Sevrin, S.~van der Loo and S.~Van Gils,
``Properties of semichiral superfields,''
  Nucl.\ Phys.\ B {\bf 562}, 277 (1999)
  [hep-th/9905141].
  %%CITATION = HEP-TH/9905141;%%
  
        %\cite{Ivanov:1994ec}
\bibitem{Ivanov:1994ec} 
  I.~T.~Ivanov, B.~b.~Kim and M.~Rocek,
  ``Complex structures, duality and WZW models in extended superspace,''
  Phys.\ Lett.\ B {\bf 343}, 133 (1995)
  [hep-th/9406063].
  %%CITATION = HEP-TH/9406063;%%
  
  
  
   
  
 %\cite{Goteman:2009xb}
\bibitem{Goteman:2009xb} 
  M.~Goteman and U.~Lindstrom,
  ``Pseudo-hyperkahler Geometry and Generalized Kahler Geometry,''
  Lett.\ Math.\ Phys.\  {\bf 95}, 211 (2011)
  [arXiv:0903.2376 [hep-th]].
  %%CITATION = ARXIV:0903.2376;%%
  
    

  
   %\cite{Lindstrom:1994mw}
\bibitem{Lindstrom:1994mw} 
  U.~Lindstrom, I.~T.~Ivanov and M.~Rocek,
``New N=4 superfields and sigma models,''
  Phys.\ Lett.\ B {\bf 328}, 49 (1994)
  [hep-th/9401091].
  %%CITATION = HEP-TH/9401091;%%
  
   %\cite{Sevrin:2011mc}
\bibitem{Sevrin:2011mc} 
  A.~Sevrin, W.~Staessens and D.~Terryn,
  ``The Generalized Kahler geometry of N=(2,2) WZW-models,''
  JHEP {\bf 1112}, 079 (2011)
  [arXiv:1111.0551 [hep-th]].
  %%CITATION = ARXIV:1111.0551;%%

 

   \end{thebibliography}
\end{document}